\documentclass[sn-mathphys,Numbered,iicol]{sn-jnl}% Math and Physical Sciences Reference Style
\usepackage{graphicx}%
\usepackage{wrapfig}%
\usepackage{multirow}%
\usepackage{amsmath,amssymb,amsfonts}%
\usepackage{amsthm}%
\usepackage{mathrsfs}%
\usepackage[title]{appendix}%
\usepackage{xcolor}%
\usepackage{textcomp}%
\usepackage{manyfoot}%
\usepackage{booktabs}%
\usepackage{algorithm}%
\usepackage{algorithmicx}%
\usepackage{algpseudocode}%
\usepackage{listings}%
\usepackage{siunitx}%
\usepackage{array}%
\usepackage{makecell, booktabs, multirow}%
\usepackage{lineno}
\usepackage{caption}

\theoremstyle{thmstyleone}%
\theoremstyle{thmstyletwo}%

\theoremstyle{thmstylethree}%

\begin{document}

\title[Article Title]{The interplay between membrane viscosity and ligand-binding receptor kinetics in lipid bilayers}
%multicomponent chemo-mechanical modeling of lipid bilayer

\author[1]{\fnm{Chiara} \sur{Bernard}}\email{chiara.bernard@unitn.it}

\author*[2,3]{\fnm{Angelo Rosario} \sur{Carotenuto}}\email{angelorosario.carotenuto@unina.it}

\author[1,4,5]{\fnm{Nicola Maria} \sur{Pugno}}\email{nicola.pugno@unitn.it}

\author*[1,7,8,9]{\fnm{Luca} \sur{Deseri}}\email{luca.deseri@unitn.it}

\author[2,3,6]{\fnm{and Massimiliano} \sur{Fraldi}}\email{fraldi@unina.it}

\affil[1]{\orgdiv{Department of Civil, Environmental and Mechanical Engineering}, \orgname{University of Trento}, \orgaddress{\country{Italy}}}

\affil*[2]{\orgdiv{Department of Structures for Engineering and Architecture}, \orgname{University of Naples ``Federico II''}, \orgaddress{\country{Italy}}}

\affil[3]{\orgdiv{Laboratory of Integrated Mechanics and Imaging for Testing and Simulation (LIMITS)}, \orgname{University of Naples ``Federico II''}, \orgaddress{\country{Italy}}}

\affil[4]{\orgdiv{Laboratory for Bioinspired, Bionic, Nano, Meta Materials \& Mechanics}, \orgname{University of Trento}, \orgaddress{\country{Italy}}}

\affil[5]{\orgdiv{School of Engineering and Materials Science}, \orgname{Queen Mary University of London}, \orgaddress{\country{UK}}}

\affil[6]{\orgdiv{Département de Physique, LPENS}, 
\orgname{École Normale Supérieure-PSL}, \orgaddress{\country{France}}}

\affil[7]{\orgdiv{Department of Mechanical Engineering and Material Sciences, MEMS-SSoE}, \orgname{University of Pittsburgh}, \orgaddress{\country{USA}}}

\affil[8]{\orgdiv{Department of Civil and Environmental Engineering}, \orgname{Carnegie Mellon}, \orgaddress{\country{USA}}}

\affil[9]{\orgdiv{Department of Mechanical Engineering}, \orgname{Carnegie Mellon}, \orgaddress{\country{USA}}}

\abstract{Plasma membranes appear as deformable systems wherein molecules are free to move and diffuse giving rise to condensed microdomains (composed of ordered lipids, transmembrane proteins and cholesterol) surrounded by disordered lipid molecules. Such denser and thicker regions, namely lipid rafts, are important communication hubs for cells. Indeed, recent experiments revealed how the most of active signaling proteins co-localize on such domains, thereby intensifying the biochemical trafficking of substances. From a material standpoint, it is reasonable to assume the bilayer as a visco-elastic body accounting for both in-plane fluidity and elasticity. Consequently, lipid rafts contribute to membrane heterogeneity by typically exhibiting higher stiffness and viscosity and by locally altering the bilayer dynamics and proteins activity. A chemo-mechanical model of lipid bilayer coupled with interspecific dynamics among the resident species (typically transmembrane receptors and trasporters) has been recently formulated to explain and predict how proteins regulate the dynamic heterogeneity of membrane. However, the explicit inclusion of the membrane viscosity in the model was not considered. To this aim, the present work enriches the constitutive description of the bilayer by modeling its visco-elastic behavior. This is done through a strain-level dependent viscosity able to theoretically trace back the alteration of membrane fluidity experimentally observed in lipid phase transitions. This provides new insights into how the quasi-solid and fluid components of lipid membrane response interact with the evolution of resident proteins by affecting the activity of raft domains, with effects on cell mechano-signaling.} 


\maketitle

\begin{table}[h!]
\centering
\caption*{List of symbols and definitions}\label{table0}
\begin{tabular}{@{}ll@{}}
\toprule
Symbol & Physical quantity \\
\midrule
$\mathbf{u}$ & Displacement field \\
$\phi$ & Transverse membrane stretch \\
$\mathbf{F}$ & Deformation gradient \\
$\mathbf{C}$ & Cauchy-Green strain tensor \\
$\mathbf{D}$ & Symmetric strain rate \\
$\mathbf{A}$ & Generic stress/strain $2$nd order tensor \\
$\mathbf{A}_0$ & Dimensionally reduced stress/strain tensor \\
 & in the membrane mid-plane \\
$\mu \left(\mu^*\right)$ & Chemical potential in the reference \\
 & (virgin) configuration \\
$\mathbf{S} \left(\mathbf{S}^*\right)$ & Stress tensor in the reference \\
 & (virgin) configuration \\
$E$ & Elastic modulus \\
$G$ & Shear modulus \\
$\nu$ & Poisson's ratio \\
$K_r$ & Remodelling term \\
$w_i$ & Chemo-mechanical coupling parameter \\
$\epsilon, \gamma$ & Constitutive parameters of the\\
& Cahn-Hilliard species potential \\
$\mathbf{Q}_i$ & Flux vector of the $i-$th species \\
$\xi$ &  G-protein coupled receptor fraction\\
$\zeta$ &  Multidrug resistance protein fraction\\
$\alpha_{\xi}$ & Uptake function \\
$\delta_i$ & Decay rates \\
$\beta_{ij}$ & Interspecific terms \\
$p$ & Lagrangian pressure \\
$\eta$ & Viscosity function \\
$\tau$ & Strain sensitivity parameter \\
$p_0$ & Applied membrane pressure \\
\botrule
\end{tabular}
\end{table}

\section{Introduction}
Early findings assumed the eukaryotic cell membranes as a bi-dimensional assembly of lipids organized in a fluid bilayer where transmembrane proteins can laterally diffuse\cite{singer1972fluid}. Lipids self-assemble in a $\sim 5 nm$ thick bilayer\cite{huang1965properties} and achieve an areal stretch of the order of $5 \%$\cite{hallett1993mechanical}. Phospholipids can move in the planar direction and, so, plasma membranes are characterized by quasi-fluid deformable surfaces that express solid-fluid-like behavior, resulting in systems wherein in-plane fluidity and elasticity may simultaneously emerge\cite{santo2022dynamical}. Such fluidity is measured through the viscosity, whose available literature data are, however, highly experiment dependent, sometimes varying by orders of magnitude\cite{faizi2022vesicle}. A possible explanation for this huge variability could be that membrane surface viscosity is a macroscopic quantity modeled at scales where the bilayer is assumed to behave like a 2-dimensional quasi-incompressible fluid. For this reason, micro- or nano- scale measurements may not be sufficient to catch the effective continuum viscosity but, rather, the so-called "microviscosity". The latter is a local quantity influenced by the environment\cite{nagao2021relationship}. Membrane fluidity is therefore associated with the high molecular mobility inside the lipid bilayer, enabling for a lateral diffusion of the embedded proteins\cite{cicuta2007diffusion}. Hence, viscosity results to be measured through the estimation of lipid diffusion coefficient\cite{faizi2022vesicle}. It is indeed confirmed that the ligand-binding of receptors --as for example the G-Protein Coulped Receptors (GPCRs)-- requires the presence of molecules that are able to move within the membrane\cite{irannejad2014gpcr}. In this regard, it has been established the difference, in terms of viscosity, among the resistance to flow under an applied shear stress and the capability of molecules to move and diffuse inside the membrane\cite{espinosa2011shear}. In the latter, it has been demonstrated that high diffusion mobility could be linked to a finite macroscopic shear viscosity, however discussing many cases of gel-phase of single saturated phospholipids or solid ceramide lipids that are able to pack themselves into a solid structure with high shear stiffness and viscosity. Quantitative stability analyses of viscoelastic lipid bilayers with properties deduced by\cite{espinosa2011shear}, have been provided in\cite{deseri2016fractional}. Furthermore, in complex bio-membranes gel domains may coexist with fluid ones, thus promoting regions with vastly distinct viscosities\cite{gohrbandt2022low}. Actually, evidences show that the mammalian cell membrane has a time-varying force response as nonlinear function of strain, so behaving as a visco-elastic or non-Newtonian fluid\cite{crawford1987viscoelastic}. Related to this phenomenology, one can recall that lipid bilayers undergo various stages at which they may experience area expansion, thereby responding with compression and shear moduli\cite{espinosa2011shear}. Such a variation in the local mechanical properties seems to be responsible for the majority of cellular processes\cite{diz2013use}. 

Several experimental strategies have been used to quantify the dynamical visco-elasticity of lipid systems\cite{choi2011active,kim2011interfacial}. Recently, AFM measurements were performed to capture both the elastic and viscous properties of lipid systems that resulted to affect the propagation or attenuation of mechano-signaling across the cell membrane\cite{al2018multifrequency}. Also, high frequency experiments, modeled through a continuum mechanical theory, revealed that the plasma membrane displays a visco-elastic behavior\cite{yu2023compressible}. In particular, it has been estimated that the cell surface responds like an elastic material on short time scales of around $1s$, while exhibiting properties of a viscous body on longer time scales $\sim 10-100s$\cite{lamparter2020cellular}. Bulk membrane viscosity and transverse stiffness are therefore correlated but also influenced by lipid packing density\cite{renne2023membrane}. 

Modulation of membrane behavior has been demonstrated to be fundamental in various diseases\cite{gleason1991excess,nadiv1994elevated,osterode1996nutritional,koike1998decreased,zubenko1999platelet}. For instance, it is indeed confirmed that changes in membrane viscosity influence the evolution of the metastatic progression of cancerous cells\cite{de1977microviscosity,adeniba2020simultaneous}. In\cite{lu2020characterization} it is shown that the latter are softer than healthy cells and that they are also characterized by a more fluid membrane. For these reasons, the measure of membrane visco-elasticity leads to the possibility of discriminating between normal and cancerous cells through the application of multi-frequency vibrations\cite{yu2023compressible}.

Lipid rafts have been demonstrated to be involved in cardiovascular signaling as determinant regulators of vascular endothelial and smooth muscle cells, and in particular in signal transduction across the plasma membrane, of primary importance to many functional activities. At present, little is known about the specific role of lipid rafts in cardiac function and dysfunction, increasing attention focusing on their contribution to the pathogenesis of several structural and functional processes including cardiac hypertrophy and heart failure, as well as atherosclerosis, ischemic injury and different myocardial functions\cite{das2009lipid}. Lipid rafts in cardiomyocyte membranes are enriched in signaling molecules and ion channel regulatory proteins, therefore contributing to calcium handling and Ca2+ entry that control excitation-contraction of heart muscle cells. Thus, they can actively participate in differential cardiomyocyte ion channel targeting and regulation\cite{das2009lipid,maguy2006involvement}. 

Ordered microdomains result fundamental to stabilize signal transduction activities required for angiogenesis. In fact, it has been observed that VEGF receptor-2 (VEGFR-2), which stimulates angiogenic signaling, co-localizes with lipid rafts to regulate its activation. Also, long-term VEGFR2 relocation closely depends on lipid raft integrity, disruption of lipid rafts directly causing receptors' depletion and inefficacy. In this sense, therapeutic strategies are more and more oriented towards the possible modulation of lipid rafts to control cells' sensitivity to VEGF expression\cite{zabroski2021lipid,ravelli2015beta3}.
Also, GPCRs have a primary influence in cardiac remodeling. Activation of epidermal growth factor receptors is in fact mediated by a large repertoire of GPCRs in the heart, and promotes cardiomyocyte survival, thus suggesting innovative therapeutic scenarios based on their targeting\cite{insel2009membrane,grisanti2017cardiac}.

Despite available pure mechanical descriptions of the lipid bilayers\cite{deseri2013stretching,maleki2013kinematics} or purely diffusive approaches where the influence of micro-mechanical stimuli is neglected\cite{garcke2016coupled}, there is still no modeling approach that takes into account the synergistic influence of membrane viscosity on transmembrane proteins activation and mobility and/or viceversa the role of proteins and lipids in membrane fluidity. Actually, it is well known that physical and chemical events act together to form the complexity of processes responsible for cell functions\cite{janmey2006biophysical}. Therefore, a multiphysics analysis becomes manifest to provide new insights into the very complex world of plasma membranes. In this regard, mathematical production provided in \textit{Carotenuto et al.}\cite{CAROTENUTO2020103974} confirmed the common knowledge that active receptors prefer to cluster on the so-called \textit{lipid rafts} --wherein high cholesterol concentration increases bilayer rigidity\cite{niemela2007assessing}-- through a chemo-mechanical coupled model. In\cite{CAROTENUTO2020103974}, the model was regulated by the coupling of the membrane remodeling and its energetics dependent on the active proteins involved in the system, i.e. $\beta2-$adrenergic receptors. Moreover, recent findings\cite{bernard2023} highlighted the effects produced by the receptors and transporters on raft formation and coalescence through Cahn-Hilliard-type dynamics in a two-dimensional hyper-elastic framework. 

Neverthless, as aforementioned the lipid bilayer is characterized by viscous properties and so, in order to obtain a more faithful solid-liquid description of this kind of system, a visco-hyperelastic model should be considered. This may provide an explicit interaction between the characteristic time evolution of the populations of transmembrane proteins and the relaxation time of the lipid bilayer. This is because, at the microscopic level, single protein re-arrangement and configurational changes are known to occur within milliseconds and are likely to locally produce elastic pressures at the membrane-protein interfaces\cite{latorraca2017gpcr,hilger2018structure}. This can be extended at the population level through the presented continuum approaches, in which the dynamics of entire protein clusters is followed in response to the ligand time-varying precipitation stimulus. The morpho-elastic reconfiguration of the membrane thus can produce maps of heterogeneous stress and deformation that could project at the continuum scale the instantaneous packing of lipids and protein activation occurring within the ordered phase.

All this considered, the aim of the present study is to enrich well-grounded hyper-elastic models\cite{bavi2014biophysical,CAROTENUTO2020103974,carotenuto2021multiscale,mahata2022computational} of cell membranes by incorporating a material viscous component in the constitutive model. This provides an explicit interaction between the characteristic time evolution of the population of transmembrane proteins and the relaxation time of the lipid bilayer, by so calling into play a possible competition between the pseudo-viscous and the characteristic viscous terms. 

\section{Chemo-Mechanical characterization of the membrane behavior}\label{sec2}
It is well established that the plasma membrane undergoes a thickness change due to an ordered-disordered phase transition occurring at the lipid scale. This thickness variation is mainly caused by the lipid re-arrangement that, in assuming an ordered configuration, have straightened tails and appear tightly packed together as it occurs in functional micro-domains of the lipid membrane denoted as raft phase\cite{simons1997functional}. Several approaches have been adopted to analyze the mechanical behavior of membrane systems when experience phase transition based on either molecular dynamics simulations or, at the continuum scale, phase separation and elasticity models\cite{uline2012phase,gauthier2012mechanical,le2019plasma,carotenuto2023towards}. Recently, a nonlinear hyperelastic response of the plasma membrane has been used to build up a fully-coupled framework describing the membrane's macroscopic remodeling and functional reorganization as regulated by the leading biochemical events occurring among interacting protein species in forming lipid raft domains\cite{CAROTENUTO2020103974}. In the subsequent work by \textit{Bernard et al.}\cite{bernard2023}, this evolutionary approach has been further enriched by Cahn-Hilliard energetics and kinetics for the involved species, thereby accounting for rafts nucleation and coalescence. The time-varying nature of the involved biological species associated to configurational remodeling terms gave to the system a pseudo-visco-elastic nature (with eventual dissipation), the rate of the internal species kindling a viscous-type (chemical) stress. However, in\cite{bernard2023} the explicit role of intrinsic visco-elasticity of the lipid membrane and the possible influence of the fluid component of the bilayer on raft development was not considered. To this purpose, we here analyze a two-dimensional system capable to experience a lipid phase separation and manifest raft coarsening within a visco-elastic environment. The whole phenomenon will be the result of the coupling between the conformational remodeling guided by the presence of the active protein species and the energetics of the membrane. In particular, the elastic part of the membrane response --in line with well-established literature\cite{evans1973new,evans1973newb,skalak1973strain}-- is modeled by assuming a neo-Hookean type behavior\cite{bernard2023}, by neglecting for now the spontaneous trends of lipids to reorganize themselves in co-existing phases (this can be accounted for not convex energy terms\cite{deseri2008derivation}). 
At the molecular scale, the activation of a single transmembrane protein within the lipid environment provokes a re-arrangement of its sub-units, which induces a stress in the surrounding membrane in the form of an in-plane pressure. This, inevitably, calls into play the adaptation of the neighboring lipids. In the absence of any viscous component, the adaptation of the lipid membrane is entirely dictated by the dynamics of the protein populations. In this sense, at the macroscopic scale the overall deformation and morphological remodeling of the lipid membrane is seen as the averaged result of the overall behavior of protein densities. The latter will pass to their active state asynchronously by introducing delays and by exchanging (positive or negative) chemical feedbacks. These give rise to more complex spatial and temporal patterns of the membrane heterogeneity. Noteworthy, the characteristic times of the membrane evolution do not simply follow the activation times of single units (of the order of few milliseconds). Rather, instead ensue the collective dynamics of active resident proteins and their progressive recruitment. Indeed, lipid and proteins' clusters have a much larger life-span (from seconds to several minutes\cite{douglass2005single,gaus2003visualizing,grassi2020lipid}). In this sense, the micro- and macro- scopic scales of the ordered macro-islands could potentially describe multi-scale kinematics in a cascade manner.  
Through the above described mechanisms, in\cite{bernard2023} an interspecific protein dynamics, enriched with a Cahn-Hilliard energetics and kinetics phenomena, has been adopted to successfully trace back the complex spatio-temporal adaptation of the membrane. Of course, the chemo-mechanical coupling becomes absolutely crucial to theoretically explain how protein density dynamics affects the structural remodeling of the membrane, leading to the nucleation of raft domains. The heterogeneity noticed in lipid bilayers has to be indeed addressed to the coexistence of disordered and ordered lipid phases\cite{hammond2005crosslinking}. To this end, well-grounded observations show the formation of zones with different concentration levels\cite{elson2010phase}. In particular, regions with high concentration of proteins have been recognized in lipid rafts\cite{simons2000lipid}, where the clustering phenomena give rise to the initiation of most of cellular processes\cite{brown1998structure,edidin2003state,chazal2003virus}. For this reason, the introduction of a phase-separation diffusive model able to predict coalescence of different species becomes apparent. Within this framework, the Cahn-Hilliard equation is typically used to describe two-phase separation problems\cite{heberle2011phase,cherfils2014generalized,duda2021coupled} that are mathematically described by a diffusion equation for the species concentration\cite{chen2002phase}. In this respect, the theoretical model proposed in\cite{bernard2023} described the evolution of protein species through Cahn-Hilliard-like energetics and kinetics wherein reaction interspecific terms account for the mutual influence among protein populations, i.e. the above mentioned GPCRs and their antagonist the Multidrug Resistance Proteins (MRPs), while non-local species momenta are enriched by strain-dependent morphotaxis terms. The latter enable the movement of protein species along the gradients of lipid order distribution, so promoting the tendency of signaling proteins to reside on raft domains by favoring spatial co-localization of such species on raft islands.
When the viscous component of the membrane is introduced and a visco-elastic behavior of the membrane is considered, the above described dynamics can be altered by the direct competition between both the characteristic adaptation and the intrinsic bilayer relaxation times. Indeed, it is expected that viscosity may affect the membrane deformation triggered by proteins through creep-associated effects in raft emergence, thus so influencing its chemical stability and persistence. On the other hand, stress relaxation phenomena could occur as well by redistributing internal stresses with effect on the residual stress-induced stiffness and membrane tension. However, rough estimations of the visco-elastic and lipid raft characteristic times --respectively of microseconds and tens of seconds-- would suggest that these phenomena would minimally concur together in determining the structural re-organization of the membrane. More important effects could be rather produced by the synergy of protein dynamics with nonlinear deformations and viscous response, which could lead instead to more significant changes into the material remodeling of membrane properties. This would meet some experimental evidences showing that rafts are highly viscous and stiff zones of the membrane. To do this, in what follows we present the governing equations of the coupled model within a visco-elastic framework. This will enable to investigate how membrane fluidity is influenced by the dynamical re-organization. In particular, we will initially consider the effects of a constant (i.e. linear) viscous term on raft persistence. While afterwords a strain-level dependent viscosity will be considered to explore if the increase of viscosity of heterogeneous lipid membranes plays a key influence on co-evolving with lipid rafts.

\subsection{Uploading visco-elasticity in the coupled chemo-mechanical model}
The lipid bilayer can be assumed as a two-dimensional quasi-incompressible hyperelastic thin body, wherein areal and thickness stretches locally vary with the corresponding changes of the lipid order\cite{evans1973new,evans1973newb,skalak1973strain}. Herein, the membrane is assumed flat in its natural configuration and its kinematics is supposed to be confined in the class of normal preserving deformations (see e.g.\cite{deseri2008derivation,deseri2013stretching,zurlo2006material}). The natural configuration of the membrane $\mathcal{B}_0$ is partitioned in a two-dimensional domain $\mathbf{x}=\mathit{x}\textbf{e}_1+\mathit{y}\textbf{e}_2$ and the thickness $z$. Hence, the material particles $\boldsymbol{\mathit{x}}\,\in\, \mathcal{B}_0$ are described as $\boldsymbol{\mathit{x}} = \mathbf{x} + z\mathbf{e_3}$, at time $t$. Accordingly, the displacement field characterizing the kinematics of the membrane can be written as follows:
\begin{equation}
\resizebox{.47\textwidth}{!}{
$\mathbf{u}(x,y,z,t)$$=[u_1(x,y,t),u_2(x,y,t),(\phi\left(x,y,t\right)-1)z]$,
}
\label{displacement}
\end{equation}
where the function $\phi \left(x,y,t\right)$ represents the thickness stretch in the direction $\mathbf{e}_3$, at time $t$. The displacement \eqref{displacement} yields the deformation gradient to which the chosen strain measures, as well as strain rates, can be readily associated:
\begin{align}\label{defs}
    & \mathbf{F}= \mathbf{I}+ \nabla \mathbf{u}, \quad \mathbf{B}= \mathbf{F} \mathbf{F}^T, \quad \mathbf{C}= \mathbf{F}^T \mathbf{F}, \nonumber \\ & \mathbf{D}= \frac{1}{2} \left(\mathbf{\dot{F}} \mathbf{F}^{-1}+ \mathbf{F}^{-T} \mathbf{\dot{F}}^T \right), \quad \dot{\mathbf{C}}=2\mathbf{F}^T \mathbf{D} \mathbf{F}.
\end{align}
By restricting the problem to the mid-plane of the membrane (see e.g.\cite{deseri2008derivation,deseri2013stretching,zurlo2006material}) and by accounting for a volumetric incompressibility constraint restricted to such mid-plane, the determinant of $\mathbf{F}$ at $z=0$ reads:
\begin{equation}
    J = J_0 \phi = 1,
\end{equation}
where $\phi\left(x,y,t\right)=\frac{1}{J_0}$, and $J_0$ denotes the areal stretch in the membrane plane, i.e. $J_0=\det\,\mathbf{F}_0$ with $\mathbf{F}_0$ defined as the dimensional reduction of $\mathbf{F}$ on the membrane mid-plane, i.e. $\mathbf{F}_0=\sum_{\alpha,\beta=1}^2 \left(\hat{\delta}_{\alpha\beta}+ \partial u_{\alpha}/\partial x_{\beta}\right)\textbf{e}_{\alpha}\otimes\textbf{e}_{\beta}$, where $\hat{\delta}_{\alpha\beta}$ is the Kronecker delta.
Incompressibility on the mid-plane also implies that $tr(\mathbf{D})=0$, once the trace is restricted to operate on $\mathbf{D}$ in such a plane. 

Following\cite{bernard2023}, the energetics of the system is assumed to be governed by the Helmholtz-free energy density $\mathcal{W}\left(\mathbf{F},n_i, \nabla n_i, \phi\right)$, where $n_i$ is the concentration of the i-th active species. Hence, by considering an additive decomposition of such energy, the contributions given by the potential associated with the hyperelastic energy of the membrane and the one related to the transmembrane proteins are introduced:
\begin{equation}
    \label{ene}\mathcal{W}=\mathcal{W}_{hyp}\left(\mathbf{F}\right) + \mathcal{W}_{n_i}\left(n_i, \nabla n_i, \phi\right).
\end{equation}
Herein, the contribution $\mathcal{W}_{n_i}$ contains a coupling term that explicitly depends on the out-of-plane stretch $\phi$, accounting for the influence that changes in species concentration have on membrane deformation and vice-versa. In fact, protein re-organization at the micro-level exerts work on the surrounding membrane, thus calling into play the bilayer deformation and stress. On this account, besides an intrinsic species-dependent energy density, $\Psi_{n_i}$, the potential $\mathcal{W}_{n_i}$ provides the coupling term due to the above mentioned interaction which reads as follows:
\begin{equation}\label{ene2}    \mathcal{W}_{n_i}\left(n_i, \nabla n_i, \phi\right) = \Psi_{n_i} - w_i\left(n_i-n_i^0\right)\left(\phi-1\right).
\end{equation}
Here $w_i$ is a coupling parameter connected to the exchange of mechanical work between activating proteins and membrane: such $w_i$ directly emerges from the sub-macroscopic scale as shown in\cite{CAROTENUTO2020103974}. As discussed above, the energy contribution $\Psi_{n_i}$ is actually given in terms of the Ginzburg-Landau phase separation energy\cite{gurtin1996generalized}:
\begin{equation}
    \Psi_{n_i} =\frac{1}{4\epsilon}n_i^2(1-n_i)^2+\frac{\gamma}{2}\left | \nabla \left(n_i-n_i^0\right)\right |^2,
    \label{psi}
\end{equation}
defining the coefficients $\epsilon$, $\gamma$ $>0$, and the gradient term $\nabla \left(n_i-n_i^0\right)$ so written to ensure thermodynamic consistency\cite{bernard2023}. More in detail, in relation \eqref{psi} a double-well potential is assumed to model the energy contribution of each species in passing from the inactive to the active state. This is done by deriving conditions for chemical equilibrium that could explicitly, although phenomenologically, take into account the effect of the fundamental mechanical coupling (i.e. the second term of \eqref{ene2}), by so modifying the energetic convenience of the system. Indeed, the cell membrane undergoes shape deformations in terms of phase transition between states separated by energy barriers. 

\begin{figure}[H]
    \centering
    \includegraphics[width=.5\textwidth]{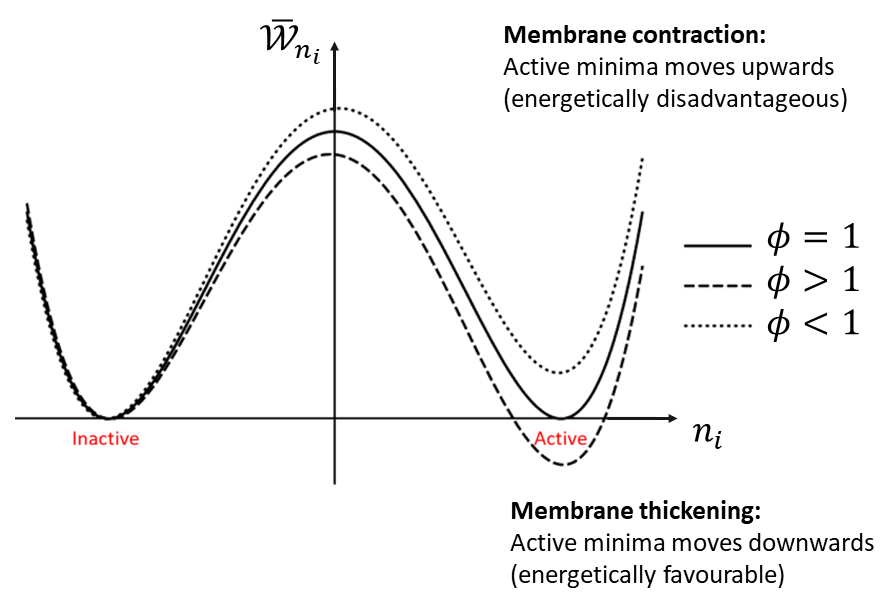}
    \caption{Qualitative influence of the membrane stretch $\phi$ on the equilibria of the double well coupled potential when a generic homogeneous density fractions is considered, i.e. $\overline{\mathcal{W}}_{n_i}=\mathcal{W} \left(n_i,0,\phi\right)$}\label{dwfig}
\end{figure}

The energy landscape of lipid membranes --and biphasic systems in general-- is modeled by a parameterized double-well potential characterized by two \textit{fixed} degenerate minima standing for the coexistence of such phases\cite{doi:10.1146/annurev-cellbio-100913-013325}. In the case of the proposed model, in presence of a varying mechanical micro-environment, the membrane mechanical state directly influences the chemical activation of the protein species. More in detail, given that in a classical double-well potential the two minima uniquely identify the active/inactive state of the proteins in a completely symmetric way, the presence of the stretch-dependent coupling term here alters such symmetry. This occurs by moving the position of the minima and so determining a non-symmetric and variable convenience of certain protein species to be in their active or inactive state on the base of the surrounding conditions. This constitutes an important mechano-signaling pathway contributing to co-localization. In fact, when the transverse stretch $\phi>1$ the coupling term makes the active state more energetically favorable with respect to the inactive one. Viceversa, as the membrane is thinning (i.e. $0<\phi<1$) the disordered state results to be more energetically convenient (see Figure \ref{dwfig}). 
In this present paper, in order to characterize the elastic part of the bilayer response, a standard incompressible neo-Hookean strain energy\cite{evans1973new,evans1973newb,bernard2023} is considered:
\begin{equation}
    \mathcal{W}_{hyp} \left( \mathbf{F} \right)= \frac{G}{2} \left(I_1-3\right) - p\,(J-1), \label{nhene}
\end{equation}
where $I_1=tr\left(\mathbf{F}^T\mathbf{F}\right)$ is the first invariant of the Cauchy-Green strain tensor and $G=E/(2(1+ \nu))$ is the tangent shear modulus with the Poisson's ratio $\nu$ approaching $0.5$ due to the incompressibility constraint, and $p$ is the associated lagrangian pressure. Consistency with linear elasticity, suggests a finite value of the elastic modulus $G$, as these two material constants are connected to each other through well-established Lam\'e relations. This is done coherent with evidence arising while observing that lipid bilayers may possess rigidity and elastic compressibility\cite{espinosa2011shear}. In fact, as reported in \textit{Espinosa et al.}\cite{espinosa2011shear}, biological membranes --for which fluidity is associated to the high molecular mobility inside the lipid bilayer enabling for a lateral diffusion of the embedded proteins-- also can account for a nonzero shear modulus as structural intrinsic property needed for biological functions.

\begin{figure*}
    \centering
    \includegraphics[width=1\textwidth]{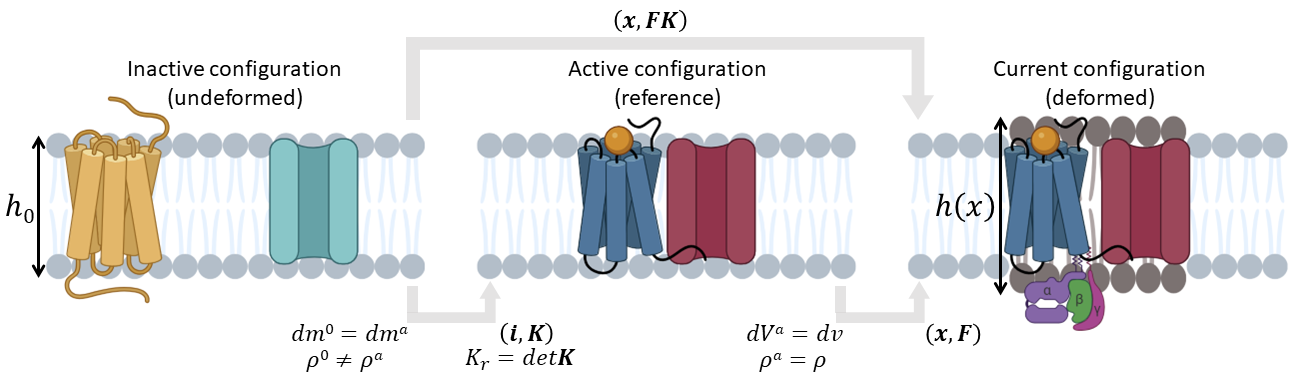}
    \caption{Active species conformational changes induce the remodeling of the lipid membrane where rafts are formed. This process is modeled through the theory of \textit{Structured Deformations}\cite{deseri2003toward,deseri2010submacroscopically,deseri2019elasticity,deseri2015stable,palumbo2018disarrangements}, a multiscale geometric framework that allows for tracing back sub-macroscopic changes in combination with classical macroscopic deformation between the active reference and the current deformed state. In the model, an inactive (undeformed) configuration is first mapped onto a geometrically identical configuration in which transmembrane proteins pass to their active state, this being characterized by the conformational jacobian $K_r$ (standing for the change in volume induced by disarrangements that are here caused by the submacroscopic remodeling). Material points in the active (reference) configuration are then mapped onto the current (deformed) one by means of the pair $\left(\textbf{x},\textbf{F}\right)$ representing the classical motion/deformation path. Here $\textbf{F} =\nabla \textbf{y}\left(\textbf{X}\right)$, and $\textbf{x} = \textbf{y}\left(\textbf{X}\right)$, where $\textbf{X}$ is a material point in the active configuration and $\textbf{y}$ represents the macroscopic deformation of the body.}\label{scheme}
\end{figure*}

Moreover, in the light of thermodynamics, as in\cite{bernard2023} it is possible to introduce specific constitutive assumptions upon which one can evaluate the stresses and the chemical potentials associated to each protein species in the presence of the chemo-mechanical coupling. 
In doing this, it is assumed that the kinematics of the remodeling membrane provides a multiple configuration path, in which the membrane is first hypothesized to undergo a geometry-preserving activation step (see Fig. \ref{scheme}).
There, part of the proteins pass to the active state by experiencing conformational switches at the sub-macroscopic scale\cite{CAROTENUTO2020103974}. At the macro-scale, this virgin-to-active state can be attained through a jacobian remodeling term, say $K_r$, derived in the framework of Structured Deformations\cite{del1993structured,deseri2003toward,deseri2010submacroscopically,deseri2019elasticity,deseri2015stable,palumbo2018disarrangements}. More in detail, this remodeling is due to submacroscopic re-arrangements of lipids clusters incorporating activated receptors. Obviously, the latter activates through conformational changes of some of their transmembrane domains during ligand-binding across the membrane. 
Thus, this depends on the amount of proteins entering the active state and it can be derived by imposing mass conservation between the virgin configuration --where material points have a virgin mass $dm^0=\rho^0\,dV^0$-- and the active (macroscopically undeformed) state, where the active mass of the material points instead read as $dm^{a}=\rho^a\, dV^a$  (see Fig. \ref{scheme}). 
Conservation of mass at the local level leads to $K_r=dV^{a}/dV^0=\rho^0/\rho^{a}$, with the densities $\rho^{(k)}$ in the heterogeneous medium being calculated as the sum of the true densities of lipids and proteins weighted by the respective fractions (see e.g.\cite{CAROTENUTO2020103974}). 
With this in mind, thermodynamical principles allow for expressing the chemical potential as: 
\begin{equation}
   \label{constm}  \mu_i^*=K_r \mu_i=K_r \left(\frac{\partial\mathcal{W}}{\partial\,n_i}-\nabla\cdot\frac{\partial\mathcal{W}}{\partial \, \nabla n_i}\right), 
\end{equation}
where, by virtue of \eqref{ene2} and \eqref{psi}, the species' chemical potentials $\mu_i$ write as follows:
\begin{equation}\label{cahn}
\resizebox{.47\textwidth}{!}{
$\mu_i=-w_i \left(\phi-1\right) + \frac{1}{2 \epsilon}n_i\left(1-n_i\right)\left(1-2n_i\right) - \nabla \cdot \gamma \nabla\left(n_i-n_i^0\right).$}
\end{equation}
On the other hand, in deriving the mechanical stresses, the Clausius-Duhem inequality leads to:
\begin{equation}\label{s1}
\left(\mathbf{S}^*-2\,K_r \frac{\partial\mathcal{W}}{\partial\mathbf{C}}\right):\frac{\dot{\mathbf{C}}}{2}\geq 0, \quad \forall\,\mathbf{C},\dot{\mathbf{C}}
\end{equation}
with $\mathbf{S}^*$ denoting the second Piola-Kirchhoff stress tensor with respect to the virgin configuration. In the present consitutively enriched model, a viscous dissipation potential $\mathcal{W}_{v}(\mathbf{C},\dot{\mathbf{C}})$ is introduced to take explicitly into account the energy dissipation due to the inherent viscosity of the membrane medium that, in the case under exam, is a pure lipid system. In this way we exclude more complex mixtures involving other structural macro-molecules such as cholesterol, whose presence in different percentages affects the membrane properties. 
Under these assumptions, the non-negative condition \eqref{s1} equates the internal dissipation such that\cite{pioletti1998viscoelastic,upadhyay2020visco}: 
\begin{equation}\label{s2}
\left(\mathbf{S}^*-2\,K_r \frac{\partial\mathcal{W}}{\partial\mathbf{C}}\right):\frac{\dot{\mathbf{C}}}{2}=K_r\,\frac{\partial\mathcal{W}_v}{\partial\dot{\mathbf{C}}}:\dot{\mathbf{C}}\geq 0, 
\end{equation}
or
\begin{equation}\label{s3}
\mathbf{S}^*=\,K_r\,\mathbf{S} =\,2 K_r \left(\frac{\partial\mathcal{W}}{\partial\mathbf{C}}+\frac{\partial\mathcal{W}_v}{\partial\dot{\mathbf{C}}}\right).
\end{equation}
This can be expressed also in terms of the Cauchy stress through a standard push-forward operation from the reference (active) to the current configuration. By considering volumetric incompressibility, one obtains:
\begin{align}\label{s4}
\boldsymbol{\sigma}^*&= \mathbf{F}\, \mathbf{S}^*\, \mathbf{F}^T= \,K_r\,\left[\frac{\partial{\mathcal{W}}}{\partial{\mathbf{F}}} \mathbf{F}^T + 2\mathbf{F}\,\frac{\partial\mathcal{W}_v}{\partial\dot{\mathbf{C}}}\,\mathbf{F}^T\right]= \nonumber \\
&=\,K_r\,\left[\frac{\partial{\mathcal{W}}}{\partial{\mathbf{F}}} \mathbf{F}^T + 2\frac{\partial\mathcal{W}_v}{\partial\mathbf{D}}\right]=K_r\,\boldsymbol{\sigma},
\end{align}
where the right-hand side of \eqref{defs} has been considered. Therefore, visco-elasticity of the membrane will depend on the specific choice of the dissipation potential. As aforementioned, the plasma membrane behaves as a visco-elastic material that experiences a vast variety of physical states with both liquid-like and solid-like behaviors\cite{espinosa2011shear}. For these reasons, viscous components could be included in a straightforward manner in order to account for such a liquid-solid description\cite{evans1976membrane}.
Herein, the stress-strain relation \eqref{s4} can be particularized through a Kelvin-Voigt-type nonlinear viscous term proportional to the rate of deformation, in order to account for rapid system variations. The Kelvin body does indeed return to its original configuration when the load, or more in general the source of deformation, is released, as typical of visco-elastic bodies\cite{meyers2008mechanical}. To this extent, it is possible to study the interplay between the characteristic relaxation time of the membrane and the protein activation dynamics in order to capture differences in lipid rafts behavior. 

Under these assumptions, the Cauchy stress tensor, with respect to the current configuration, reads as follows (see e.g.\cite{evans1976solid,pucci2010special,filograna2009simple}):
\begin{equation}
    \boldsymbol{\sigma}=\frac{\partial{\mathcal{W}}}{\partial{\mathbf{F}}} \mathbf{F}^T +2 \eta \mathbf{D}.
    \label{visco-elasticity}
\end{equation}
The viscous part of the stress is thus defined through the viscosity term $\eta>0$, which can be either constant as in the case of linear visco-elasticity or can be a function of polynomial scalar invariants involving the strain and the strain rate tensors\cite{pucci2010special,pioletti1998viscoelastic,upadhyay2020visco}. In what follows, we will focus on the effects of both possible constant viscosities as well as a strain-sensitive viscosity. In the light of this, it is worth highlighting that the particular constitutive choice in \eqref{visco-elasticity} corresponds to considering a dissipation potential of the type:
\begin{equation}
    \resizebox{.47\textwidth}{!}{$\mathcal{W}_v = \eta\left(\mathbf{B}\right) \left[\mathbf{D}:\mathbf{D}\right] = \frac{\eta\left(\mathbf{C}\right)}{4}\,\left[\dot{\mathbf{C}}:\left(\mathbf{C}^{-1}\overline{\otimes}\mathbf{C}^{-1}\right):\dot{\mathbf{C}}\right],$}
\end{equation}
where the right-hand side of \eqref{defs} has been used (the pulled-back fourth order identity tensor is defined such that $\left[\mathbf{A}\overline{\otimes}\mathbf{B}\right]_{ijhk}=A_{ih}B_{jk}$). In addition, by considering the free energy of the system \eqref{ene} involving the coupled potential \eqref{psi} and the neo-Hookean strain energy contribution \eqref{nhene} of the membrane, the Cauchy stress assumes the following expression:
\begin{equation}\label{sigma1}
\resizebox{.47\textwidth}{!}{
    $\boldsymbol{\sigma}=-p \textbf{I}+ G \textbf{F} \textbf{F}^T-w_i (n_i-n_i^0)(\textbf{e}_3\otimes\textbf{e}_3)\cdot\textbf{F}^T+2 \eta \textbf{D}$.}
\end{equation}
Under the assumption of plane stress, the out-of-plane stress component $\sigma_{33}=\mathbf{e}_3\cdot\mathbf{\sigma}\cdot\mathbf{e}_3$ vanishes thus leading to estimate the pressure $p$. By restricting the deformation gradient in the mid-plane of the membrane, one has that:
\begin{equation}\label{p1}
    p=G \phi^2-w_i \left(n_i-n_i^0\right) \phi+ 2 \eta\frac{\dot{\phi}}{\phi} .
\end{equation}
This allows to obtain the in-plane Cauchy stress $\boldsymbol{\sigma}_0$ as follows:
\begin{equation}
\resizebox{.47\textwidth}{!}{
    $\boldsymbol{\sigma}_0= G\left(\textbf{F}_0\textbf{F}_0^T-\phi^2\,\textbf{I}_0\right)+ w_i (n_i-n_i^0)\phi\,\textbf{I}_0+2 \eta\left(\textbf{D}_0-\frac{\dot{\phi}}{\phi}\,\textbf{I}_0\right)$,}
\end{equation}
in which $\textbf{I}_0$ and $\textbf{D}_0$ are respectively the in-plane identity operator and the strain rate. In order to write equilibrium with respect to the reference domain, the in-plane nominal stress tensor can be obtained through a Piola transformation as $\mathbf{P}_0= \boldsymbol{\sigma}_0 \mathbf{F}_0^{-T}$, so having:
\begin{equation}\label{1PK}
\resizebox{.47\textwidth}{!}{
$\textbf{P}_0$$=G$ ($\textbf{F}_0$-$\phi^2$$\textbf{F}_0^{-T}$)$+w_i (n_i-n_i^0)$ $\phi$$\textbf{F}_0^{-T}$$+2$ $\eta$ ($\textbf{D}_0$-$\frac{\dot{\phi}}{\phi}\textbf{I}_0)$$\textbf{F}_0^{-T}$,
}
\end{equation}
where the relation $\dot{\phi}=-\phi (\dot{\textbf{F}}_0:\textbf{F}_0^{-1})$ is employed because of incompressibility. Consequently, the pulled-back stress reads as follows:
\begin{equation}
    \textbf{P}_0^*=K_r\,\textbf{P}_0.
\end{equation}
By neglecting body forces and inertia terms, the mechanical equilibrium of the membrane reads:
\begin{equation}
    \nabla_0 \cdot \mathbf{P}_0^*=\textbf{0},
    \label{piola}
\end{equation}
with $\nabla_0$ representing the in-plane nabla operator in the virgin configuration. 

As said, the mechanical stress terms involve the co-action of resident transmembrane protein species, whose dynamics induce the rearrangement of the membrane and, in turn, its overall deformation. Therefore, the coupled system at hand must provide the presence of species-related mass balances. The generic mass balance equations for the \textit{i-}th species $\dot{n}_i$, given in terms of the species' reference flux $\mathbf{Q}_i$ and the interspecific rates $\Gamma_i$, are thus calculated according to the above attained chemical potential:
\begin{equation}
\dot{n}_i=-\nabla\cdot\mathbf{Q}_i +\Gamma_i.
\end{equation}
The flux term $\mathbf{Q}_i=-L_i \nabla \mu_i^*$ refers to the driving force $\nabla \mu_i^*$ generating species momentum in the mass balance and mediated by the scalar diffusion mobility parameter $L_i$. While, the source term $\Gamma_i$ measures chemical interactions between the two protein populations, namely GPCRs and MRPs indicated with $\xi$ and $\zeta$ respectively. Given their mutual interaction extensively explained in\cite{bernard2023}, through Volterra-Lotka-like interspecific terms, the mass conservation equations write:
\begin{equation}
\label{bal}
    \begin{cases} \dot \xi + \nabla \cdot \mathbf{Q}_{\xi} = \xi \left(\alpha_{\xi} - \delta_{\xi} - \beta_{\xi \zeta} \zeta\right) \\ \dot \zeta + \nabla \cdot \mathbf{Q}_{\zeta} = \zeta \left(- \delta_{\zeta}  + \beta_{\zeta \xi} \xi\right)
    \end{cases},
\end{equation}
where such dynamics is regulated by the decay rates $\delta_i$, the interspecific terms $\beta_{ij}$ and the activation term $\alpha_{\xi}$ that regulates the activity of GPCRs. More specifically, the uptake function $\alpha_{\xi}$ accounts for the response of the receptor to the ligand precipitation rate whose kinetics is controlled in time by a generic Gamma distribution $\gamma\left(t\right)$ and spatially by a distribution function $\iota(\mathbf{x})$. Therefore, one can write $\alpha_{\xi} = k_b Q^{-1} \iota(\mathbf{x})\gamma \left(t \right)$, where $k_b$ is defined as the binding constant, and $Q$ is the total quantity of ligand averaged over the membrane area\cite{bernard2023}.

All the values adopted for the numerical study are reported in Table \ref{table}.

\begin{table*}[h!]
\centering
\caption{Summary of the numerical values for the coefficients used in the model.}\label{table}
\begin{tabular}{@{}llll@{}}
\toprule
Coefficient & Value[Unit] & Range[Unit] & Reference \\
\midrule
$L_i$ & $7\mathrm{x} 10^{-17}[m^{2}Pa^{-1}s^{-1}]$ & $\left(10^{-20}-10^{-15}\right)[m^{2}Pa^{-1}s^{-1}]$ &\cite{CAROTENUTO2020103974,quemeneur2014shape,alenghat2013membrane,jacobson2019lateral}\\
     $k_b$ & $5.18$ & $3.89-5.7$ & \cite{bridge2018modelling,li2017lipid}  \\
     $Q$ & $2000 [pMol]$ &  & \cite{CAROTENUTO2020103974}  \\
     $\delta_{\xi}$ & $1.1 \mathrm{x} 10^{-3} [s^{-1}]$ & $\left(0.9-1.65\right)\mathrm{x} 10^{-3} [s^{-1}]$ & \cite{bridge2018modelling} \\
     $\delta_{\zeta}$ & $10^{-7} [s^{-1}]$ & $\left(10^{-8}-10^{-6}\right) [s^{-1}]$ & \cite{CAROTENUTO2020103974} \\
     $w_{\xi}$ & $5.25[MPa]$ & $\left(5-8\right)[MPa]$ & \cite{CAROTENUTO2020103974} \\
     $w_{\zeta}$ & $2.25[MPa]$ & $\left(2.17-3.5\right)[MPa]$ & \cite{CAROTENUTO2020103974} \\
     $\beta_{\xi \zeta}$ & $1.25\mathrm{x} 10^{-2}[s^{-1}]$ &  & -  \\
     $\beta_{\zeta \xi}$ & $1.28\mathrm{x} 10^{-2}[s^{-1}]$ &  & - \\
    $\xi^0$ & $10^{-1}$ &  & -  \\
     $\zeta^0$ & $10^{-2}$ &  & -  \\
     $\epsilon$ & $0.05[Pa^{-1}]$ &  & -  \\
     $\gamma$ & $0.1[Pa. \mu m^2]$ &  & -  \\
     $\eta$ & & $\left(10^{-3}-10^6\right)[Pa .s]$ - \small{fluid/gel visco-elastic systems} & \cite{faizi2021viscosity,cicuta2007diffusion,nagao2021relationship,zgorski2019surface,faizi2022vesicle, espinosa2011shear}  \\
     $\,$ & & $\left(10^{7}-10^9\right)[Pa .s]$ - \small{tough visco-elastic systems} & \cite{rand1964mechanical,katchalsky1960rheological}  \\ 
     $E$ & & $\left(2-13\right)[MPa]$ & \cite{bavi2014biophysical,carotenuto2023towards}  \\
     $\overline{\phi}$ & $1.1$ & & - \\
     $\chi$ & $50$ & & - \\
\botrule
\end{tabular}
\end{table*}

\subsection{Governing equations of the model}

Given the well-established interplay between GPCRs structural and functional organization of the cell membrane and the bilayer thickness and stress variations\cite{bernard2023}, we now present the governing equations regulating the modeled dynamics. In this sense, the mechano-biological process turns out to be governed by the balance of linear momentum in (\ref{piola}) and the time-evolution laws in (\ref{bal}) for the two protein fractions GPCRs and MRPs involved in the ligand-binding. Indeed, these species have been selected as the main families of transmembrane proteins that participate to the regulation of the membrane micro-environment. Therefore, one has the following set of coupled equations:
\begin{equation}
\label{coupling}
    \begin{cases} \nabla_0 \cdot \mathbf{P_0}^* = \textbf{0} \\ 
    \dot \xi + \nabla \cdot \mathbf{Q}_{\xi} - \xi \left(\alpha_{\xi} - \delta_{\xi} - \beta_{\xi \zeta} \zeta\right) = 0 \\ 
    \dot \zeta + \nabla \cdot \mathbf{Q}_{\zeta} - \zeta \left(- \delta_{\zeta} + \beta_{\zeta \xi} \xi\right) = 0
    \end{cases}.
\end{equation}
Numerical solutions of such system have been implemented in the software COMSOL Multiphysics\textsuperscript{\textregistered}\cite{com}, by adopting a monolithic scheme of fully coupled PDEs solved simultaneously by using a Newton nonlinear method and by discretizing the domain through a Delaunay tessellation. This by considering a circular domain $\Omega = \{(x,y)\in R^2:x^2+y^2 \le R^2 \}$ with $R=5\mu  m$, and a time span $t \in \left[0,t_{max}\right]$, where $t_{max}=1 h$\cite{bernard2023}. Provided constant initial conditions for the protein fractions $\zeta \left(x,y,0\right) = \zeta^0$ and $\xi \left(x,y,0\right) = \xi^0$, the in-plane displacements are both set with null initial values $\mathbf{u}\left(x,y,0\right)=\mathbf{0}$. Also, null species fluxes imply the boundary condition $\nabla n_i \cdot \hat{\mathbf{N}} = 0$ for the proteins and a stress-prescribed situation with a non-zero radial stress at the boundary is considered to simulate the  Laplace membrane tension due to the intra-cellular pressure. Therefore, the nominal traction in the radial direction at the outer radius writes $\mathbf{P}_0^* \cdot \hat{\mathbf{N}}= T_R\hat{\mathbf{N}}$, which can be evaluated through a prescribed outer (actual) pressure $p_o$ by imposing the equivalence $p_o\,h\,ds=T_R\,h_0\,dS^0$ that leads to $T_R=p_o(1+u_R/R)/J_0$, where $u_R$ stands for the magnitude of the in-plane displacement at the boundary.
In the following section, we will show the influence of viscous dissipation on the solid-liquid behavior of plasma membranes under different conditions able to reproduce scenarios in which membrane's morphology and mechanical adaptation lead to various situations.

\section{Results and discussion}
Within the framework of membrane visco-elasticity, we here present numerical results that permit to observe the viscosity landscape of the phase-separated domains, by focusing on possible differences in terms of raft lifespan and heterogeneity. To this aim, sensitivity analyses will be carried out to map the evolution of an initially (geometrically and materially) homogeneous membrane, by observing how raft domains and viscosity change. This will be mainly investigated as a function of the membrane's (elastic and viscous) tangent properties and initial protein distributions. In the light of the pivotal role of mechanics in the spatio-temporal dynamics of the raft-associated proteins, we analyze protein-induced adaption processes. Indeed, conformational changes of GPCR and MRP populations are capable to induce the overall remodeling of the bilayer at the membrane scale. With this in mind, in order to trigger the activation dynamics, we consider the realistic situation in which extracellular molecules randomly precipitate on the domain. This is done by assigning a random distributions to the ligand precipitation rate functions used in \eqref{bal} and by modulating the amount of precipitating ligand to induce differential receptor responses, thus orienting the membrane dynamics towards various patterns.

In numerical analyses, we start from studying the effects of a constant viscosity on the spatio-temporal behavior of the ordered phase. To then investigate more in depth the material adaptation of the bilayer in terms of the evolution of viscous properties of the rafts through a strain-sensitive viscosity term. This enrichment allows to follow the strain-induced remodeling of the lipid phase. In particular, this is done by meeting wide literature evidences demonstrating that viscosity of ordered clusters tends to increase as the phase order increases\cite{sakuma2020viscosity}. 
Starting from the initial Newtonian hypothesis, sensitivity analyses are carried out by varying the viscosity over a range compatible with literature data. In this respect, surface shear viscosity seems to exhibit a large variability depending on the particular composition of the mixed lipid system, on the specific conditions in which tests are performed as well as on the adopted experimental methods. Typical values of tangent viscosity for the most of biological membranes result of the order of $10^{-3}-10^2 Pa.s$\cite{espinosa2011shear,deseri2016fractional,den2007intermonolayer,faizi2021viscosity,faizi2022vesicle}. Fewer cases were found to instead exhibit significantly higher tangent viscosities ranges of $10^5-10^6 Pa.s$\cite{espinosa2011shear,faizi2021viscosity}, up to peaking to unusual values $10^9 Pa.s$ in case of the so-called \textit{tough} visco-elastic systems\cite{rand1964mechanical,katchalsky1960rheological}.
However, it is worth highlighting that these experimental observations report significant differences when cholesterol is introduced in the mixed lipid systems. In particular, cholesterol highly affects the stiffening and the viscosity increase of the membranes and it has a direct impact on raft stabilization as well\cite{faizi2021viscosity,nipper2008characterization,doole2022cholesterol}. In the present model, we limit our analyses to pure and mixed lipid systems, for now excluding the explicit modeling of cholesterol as a structural component of the membrane medium, which could be instead taken into account through the suitable determination of homogenized material properties depending on the extent of cholesterol fraction.

\subsection{Insights on the influence of tangent stiffness and viscosity on membrane remodeling from a Newtonian model}

First, we assume the simplest case with a constant viscosity term $\eta$, whose range of variability is reported in Table \ref{table}. This is considered as a mean shear viscosity, evaluated on the whole membrane, that does not take into account the fluidic variation in phase transitions. When $\eta$ is a constant, given the wide range of viscosity values, outcomes have been organized and presented by referring to two classes of visco-elastic responses, denoted as the \textit{weak} and the \textit{tough} visco-elastic systems. The former case indicates Newtonian viscosities lying in the wide range $10^{-3}-10^5 Pa.s$, which characterizes most of the biological membranes encountered throughout the literature. Their behavior varies from that one of a low viscosity fluid to that of a visco-elastic gel. In such a situation, linear visco-elasticity results to minimally interfere with the chemo-mechanical activity of the membrane and the overall dynamics almost entirely protein-dominated. The most important differences are indeed appraised by varying the initial stiffness of the membrane, which really does affect the coupling. The tangent Young's modulus is assumed to vary so that the membrane can undergo different configurations in the solid-fluid transition. Indeed, the stiffness of the environment mediates the mechanical work performed by proteins on the lipid medium. 

\begin{figure*}[h!]
    \centering
    {\includegraphics[width=1\textwidth]{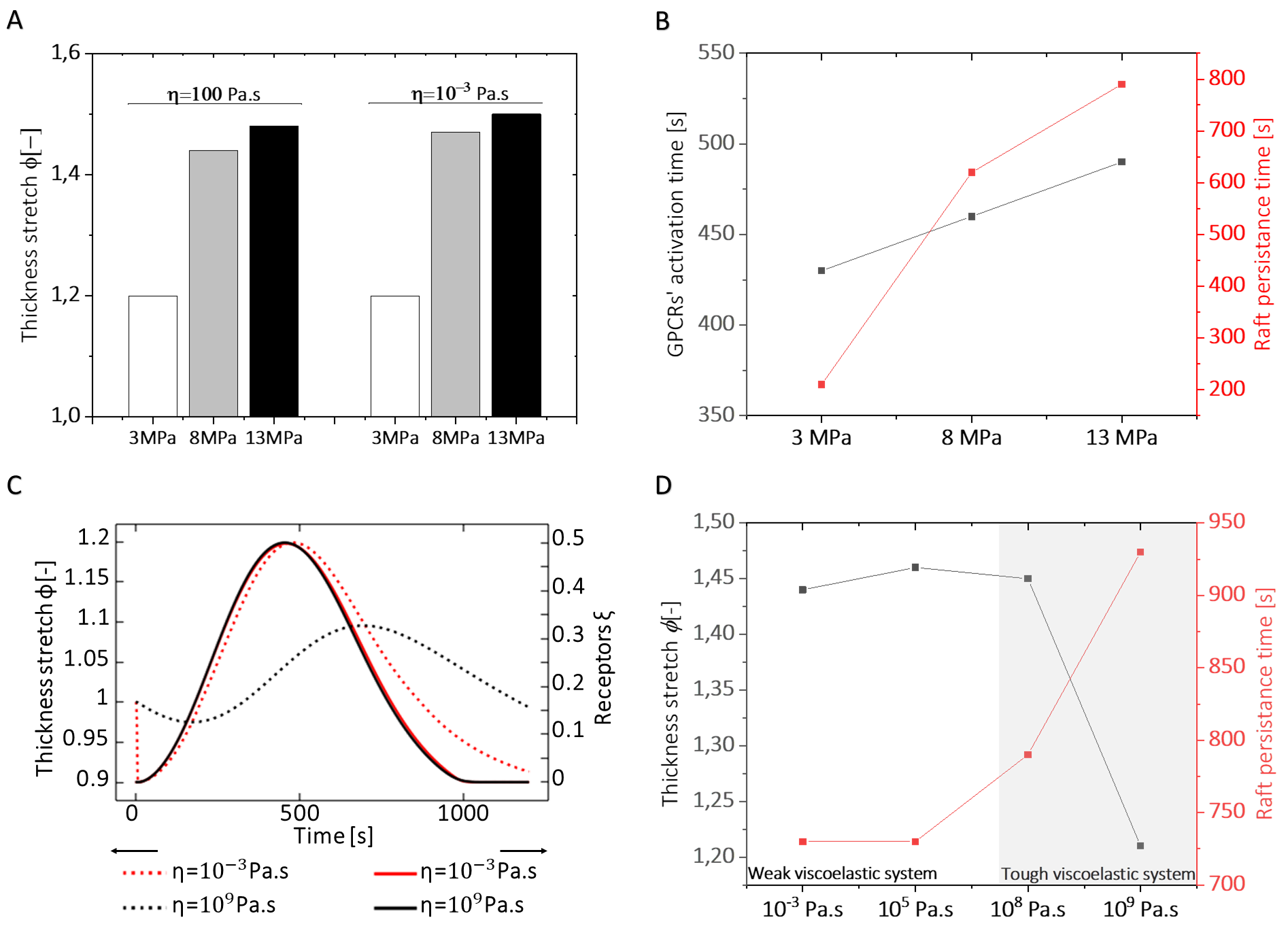}}
    \caption{Lipid membrane response to elastic and dissipative variations. \textbf{A}: Thickness stretch $\phi$ measured at constant viscosities with varying Young's modulus. Viscosity variation does not significantly affect the out-of-plane deformation that is instead influenced by changing in membrane rigidity. \textbf{B}: At fixed $\eta=100 Pa.s$, membrane undergoing deformability and rigidity results in changing the activity of GPCRs and the formation of rafts domains.
    \textbf{C}: Influence of weak and tough viscosities on the morphological re-organization of the membrane in response to analogous GPCRs activity.
    \textbf{D}: Thickness stretch and raft domains persistance measured for weak and tough visco-elastic systems. Highly viscous system leads to variations in membrane remodeling.}
    \label{istogrammi}
\end{figure*}

By considering as representative, and most frequent, cases for the weak visco-elastic systems the values $\eta=\eta_1=100 Pa.s$ and $\eta=\eta_2=10^{-3} Pa.s$, Fig. \ref{istogrammi}\hyperref[istogrammi]{A} shows that the thickness stretch is mostly determined by variations in the elastic part rather than the dissipative one. It indeed increases at higher Young's moduli, though it does not significantly change when different viscosity values are employed. Coherently with literature findings\cite{yuan2002size}, the out-of-plane deformation results to be in a range of about $20-50\%$. It is worth to note that the coupling parameters $w_i$ vary proportionally with the elastic modulus by so influencing the overall membrane activity and deformability. In fact, as such coefficient translates the microscopic mechanical interaction at the protein subunit-membrane interface, it results to be proportional to the local surface tension. That inevitably involves the stiffness of the lipid medium\cite{CAROTENUTO2020103974}. 
Moreover, for the higher viscosity $\eta_1=100 Pa.s$, the influence of the elastic part results in both the activation time of the raft-associated proteins GPCRs and the persistence of $L_o$ phase in the bilayer (see Fig. \ref{istogrammi}\hyperref[istogrammi]{B}). As shown, in the case of a more deformable system, the receptor-ligand biding occurs at $t \simeq 430 s$ accompanied by a faster raft duration of about $10 s$. Stiffer membranes instead produce a slower response of GPCRs, although a larger duration of the $L_o$ domain up to a lifespan of $100 s$ is ensured. Noteworthy, these delays in the activation times of Fig. \ref{istogrammi}\hyperref[istogrammi]{B} can be produced by the competition of the viscosity with the internal protein dynamics. The latter emerges from the complex interplay of protein intrinsic rates and stiffness-associated work terms influencing their spatio-temporal evolution through the species' momentum terms. 

\begin{figure*}[ht]
    \centering
    {\includegraphics[width=.86\textwidth]{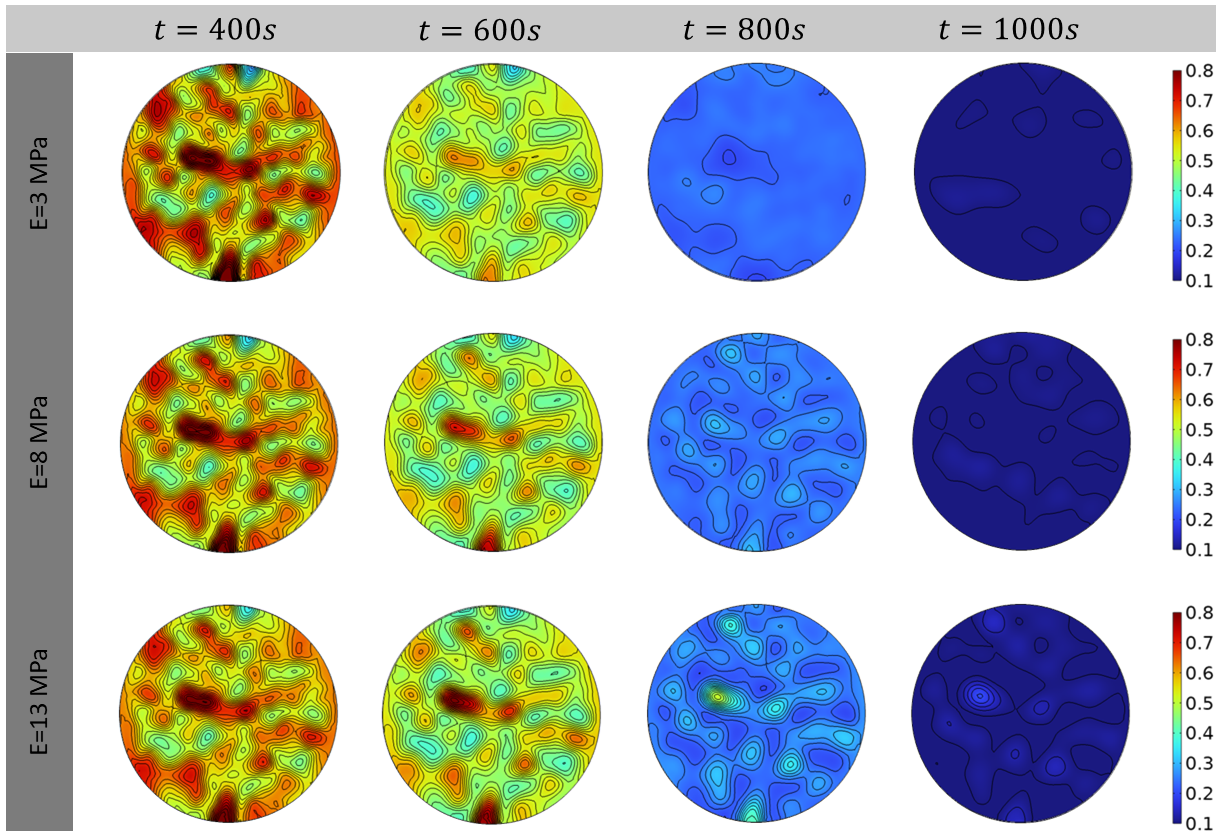}}
    \caption{Surface plots showing the active GPCRs domains in the visco-elastic system with fixed $\eta=100 Pa.s$ and varying elastic moduli. Such a variation influences membrane remodeling and configuration. It is indeed evident that a more rigid surface leads the rafts islands to be more persistent in time by reducing the lateral mobility of transmembrane proteins.}
    \label{constantviscosity}
\end{figure*}

The low influence of Newtonian viscosity \textit{de facto} suggests to adopt nonlinear viscosity models. To get more insights into the influence that a constant viscosity term can have on membrane dynamics, we carried out --at least as illustrative theoretical cases-- simulations that take in consideration the extreme situation of \textit{tough} visco-elastic membranes. This is reported to the best of Authors' knowledge in few literature works concerning the characterization of red blood cells' membranes\cite{rand1964mechanical,katchalsky1960rheological}. By thus prescribing steep values of viscosity capable to interfere with membrane dynamics, it is possible to observe a drastic change of the bilayer's morphological response to the activation of protein populations. Indeed, as shown in Fig. \ref{istogrammi}\hyperref[istogrammi]{C}, GPCRs evolve in a substantially analogous manner both in the weak and tough visco-elastic cases, since they respond to the same imposed chemical stimulus. On the other hand, in the fluid case, after the initial contraction due to the applied tension, membrane thickening grows with strong synergy and has a reduced relaxation delay following the GPCRs' decay. Conversely, in the tough system, raft emergence forms with much slower velocity. There, the extremely viscous environment highly reduces the proteins' mobility, by preventing their capability to exert mechanical work against the membrane, and by also inducing high retardation in the morphological adaptation of the plasma medium to receptors' desensitization. This is confirmed in Fig. \ref{istogrammi}\hyperref[istogrammi]{D} at different viscosities. In the fluid-gel regime, dynamics leads to co-localized and almost synchronous progression with similar morphological rearrangement, this drastically decelerating in tough visco-elastic systems with a consequent decline of the out-of-plane reconfiguration. In the light of these considerations, the latter cases demonstrate that high initial viscosity contrasts the highly dynamic and heterogeneous character of plasma membranes, by compromising the co-evolution capability. That allows the bilayer to exhibit a sufficiently reactive morphological adaptation able to favor the formation of ordered domain working as necessary sights for chemical signaling. 

Then, with reference to more common visco-elastic gel-like systems (at $\eta_1=100 Pa.s$), differences in durability can be captured in terms of prolonged protein activity in stiffer environments. In fact, as reported in Fig. \ref{constantviscosity}, variations in the persistence of receptor ligand-binding reflect the spatial organization of the bilayer in terms of raft emergence and membrane relaxation. Although the maximum activity of GPCRs occurs at slightly different times, as observable starting from $t \simeq 400 s$, the thickened $L_o$ domains decay faster in the softer membranes --being they almost extincted already at $800 s$-- while the formed GPCRs clusters are still active in membranes with a higher degree mechanical interaction. 

\subsection{Effects of strain-sensitive viscosity and evolution of membrane fluidity}

Further information can be envisaged by introducing a more complex viscous term in the model. Indeed, nonlinear effects could occur during moderate-to-large strains. In turn, this could involve non-Newtonian responses for the shear viscosity. In this way, it is possible to capture the effective fluidity of the membrane upon large strength motions\cite{espinosa2011shear}. For this reason, a strain-level dependent viscosity is assumed in a purely phenomenological fashion. This allows us to investigate situations able to theoretically confirm that the viscosity depends on membrane composition, thus it varies following ordered-disordered phase transition\cite{sakuma2020viscosity}.

\begin{figure}[ht]
    \centering
    \includegraphics[width=0.4\textwidth]{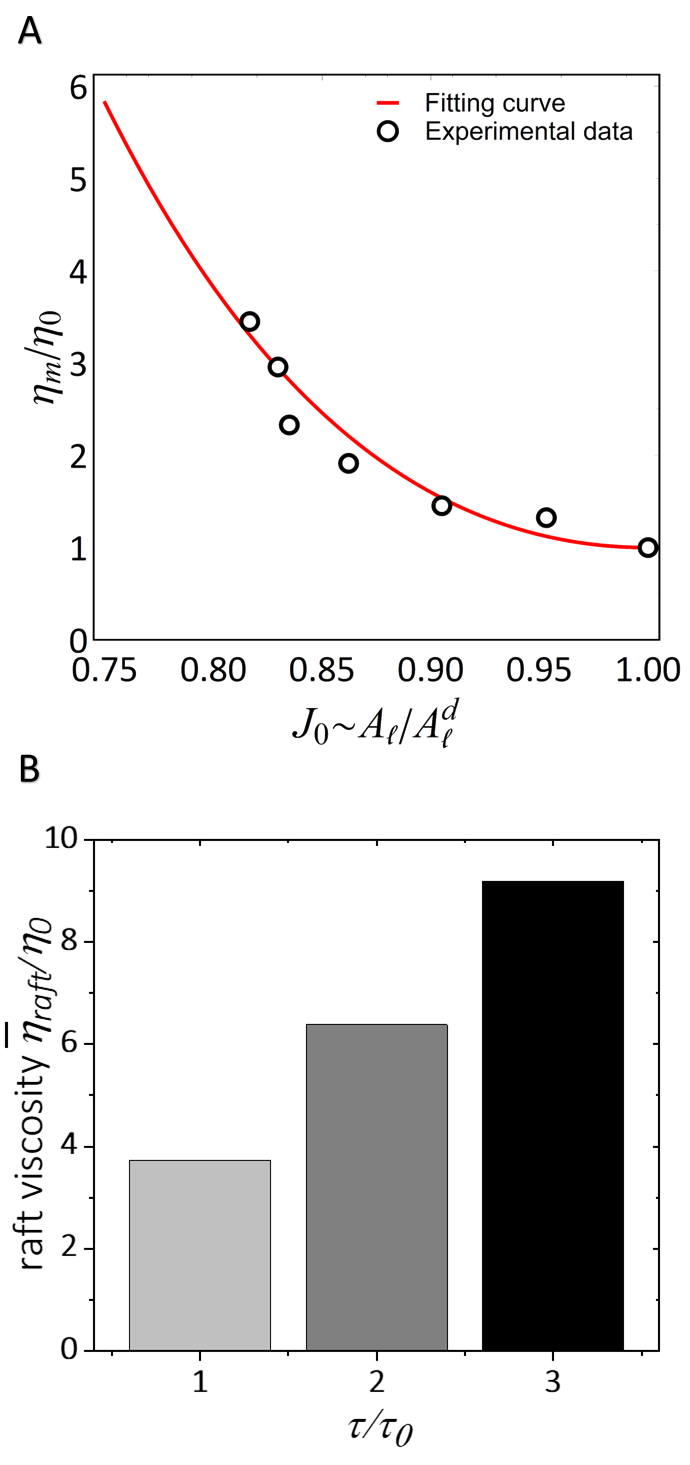}
    \caption{Fitting parameter $\tau_0$.
    \textbf{A}: Determination of the viscosity sensitivity to membrane strain. Data adopted from\cite{kelley2020scaling}.  
    \textbf{B}: Analysis of strain-induced viscosity, at maximum protein activity, for different strain sensitivity values $\tau$.}
    \label{tau}
\end{figure}

To this aim, among the possible constitutive choices and in order to introduce an essential functional variability (see e.g.\cite{pucci2010special,pioletti1998viscoelastic, upadhyay2020visco}), we assume that the viscosity term is a function of the right Cauchy-Green strain tensor through its first invariant. This is done here by means of the expression $\eta_m=\eta_0 \left[1+ \tau_0 \left(tr(\textbf{C})-3\right)\right]$. Herein, the tangent (Newtonian) viscosity $\eta_0$ has been set equal to $\eta_1$, being it compatible with the order of magnitude of the most of lipid systems. Furthermore, the coefficient $\tau_0$ is a non-dimensional parameter modulating the sensitivity to the strain. In order to determine a proper value of this latter coefficient, we exploited data in \textit{Kelley et al.}\cite{kelley2020scaling}, reporting experiments and associated scaling relationships for the viscosity of mixed lipid membranes as a function of the lipid area per unit molecule. In particular, as also shown in Fig. \ref{tau}\hyperref[istogrammi]{A} the lower is the available area per lipid the higher results the viscous term. In the present continuum approach, the area per unit lipid molecule can be put in direct correlation with the in-plane areal stretch $J_0$. To this end, by assuming a homogeneous deformation, one can fit experimental points to calibrate the proposed strain-dependent viscosity law, so deriving a reference value for the fitting parameter $\tau_0$ ($\tau_0=17.35$).
However, in order to account for the large variability of membrane fluidic properties and investigate the influence of strain sensitivity, possible variations of the parameter $\tau_0$ have been prescribed during the numerical simulations (three values proportional to $\tau_0$ have been assumed).
The proposed phenomenological law for the viscosity proposed above has been then uploaded in the coupled model in order to analyze the evolution of raft viscosity during membrane activity. In particular, the effective viscosity of raft domains has been evaluated as the tangent viscosity at the achieved strain level as 
$\overline{\eta}_{raft}={A_{raft}^{-1}}\int_{A} f\left(\phi\right) \eta_0\,K_r\,\left[1+ \tau_0 \left(tr(\textbf{C})-3\right)\right]\, dA$, with the auxiliary function $f$ defined to select raft zones as $f\left(\phi\right)=(1+\tanh(\chi\left(\phi-\overline{\phi}\right))$, while the raft area coverage results $A_{raft}=\int_{A} f\left(\phi\right)\, dA$ (see the Appendix for details on tangent viscosity).
As it can be noticed in Fig. \ref{tau}\hyperref[istogrammi]{B}, the numerical simulations show that raft viscosity intensifies from four up to ten times at the moment of maximum activity, depending on the strength of strain sensitivity. These increments are consistent with many experimental works reporting that $L_o$ phases exhibit a higher viscosity than the $L_d$ domains\cite{kelley2020scaling,sakuma2020viscosity,faizi2021viscosity,faizi2022vesicle,wu2013molecular}. Thus, this approach suggests that the adopted nonlinear viscosity can represent a proper strategy to predict the dynamic changes of membrane fluidity during order transitions. 

\begin{figure}
    \centering
    \includegraphics[width=0.45\textwidth]{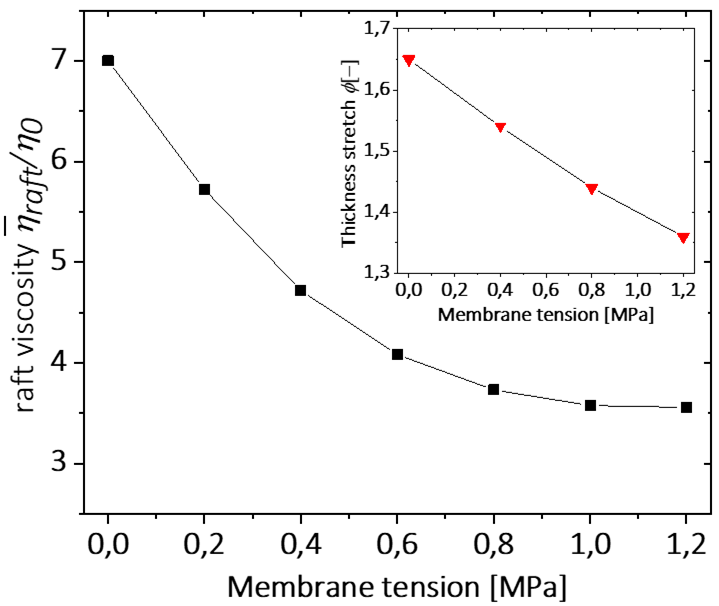}
    \caption{Membrane mechanical properties evaluated at different membrane tensions. The viscosity of the $\phi_{L_0}$ domain decreases as the pressure $p_0$ increases in the range of $0-1.2 MPa$, as well as membrane thickening, suggesting that such mechanical properties varies with the intra-cellular stimuli.}
    \label{tension}
\end{figure}

Noteworthy, the strain-dependent membrane shear viscosity can be affected by the intra-cellular tension that acts on the bilayer in both structural and dynamical properties\cite{reddy2012effect}. Therefore, we performed simulations with different pressures $p_0$ at the stress-prescribed boundary. Outcomes are shown in Fig.\ref{tension} where, according to literature findings\cite{sens2015membrane}, the membrane tension ranges from $0.1MPa$ to $1.2MPa$. Such values are consistent with the levels of intracellular pressures (Laplace's law implies that $p_0\propto p_{cell}\times R_{cell}/2h_0\simeq 10^3\,p_{cell}$, being the intracellular pressure of the order of $0.01-1$ kPa\cite{petrie2014direct}) and keep below the estimated rupture tension of $2MPa$\cite{tan2011rupture}. From Fig.\ref{tension} one can also show that, at fixed $\tau=\tau_0$, the effective raft viscosity $\overline{\eta}_{raft} / \eta_{0}$ tends to decrease as the intra-cellular pressure increases. Such behavior is reasonable with the established relationship between membrane tension and bilayer mechanical response\cite{reddy2012effect,paez2021imaging}. Indeed, increasing pressure reduces membrane thickness and works for areal expansion. It 
competes against the morpho-taxis phenomena involving membrane thickening and contrasting the tendency of transmembrane proteins to aggregate, thereby reducing the ligand-binding effectiveness and resulting in lower $L_o$ volume fraction.

It is then apparent that membrane shear viscosity varies with lipid phase order. This is due to the fact that ordered-phase islands exhibit a higher level of lipid packing compared to $L_d$ domains, by so resulting to be less polar and more viscous\cite{m2008liquid}. In particular, according to literature measurements, the $L_o$ regions seem to be characterized by a membrane viscosity higher than the one of the $L_d$ phase\cite{honerkamp2013membrane,santos2016lipid,amador2021hydrodynamic,faizi2022vesicle}. 

\begin{figure*}[ht]
    \centering
    \includegraphics[width=.75\textwidth]{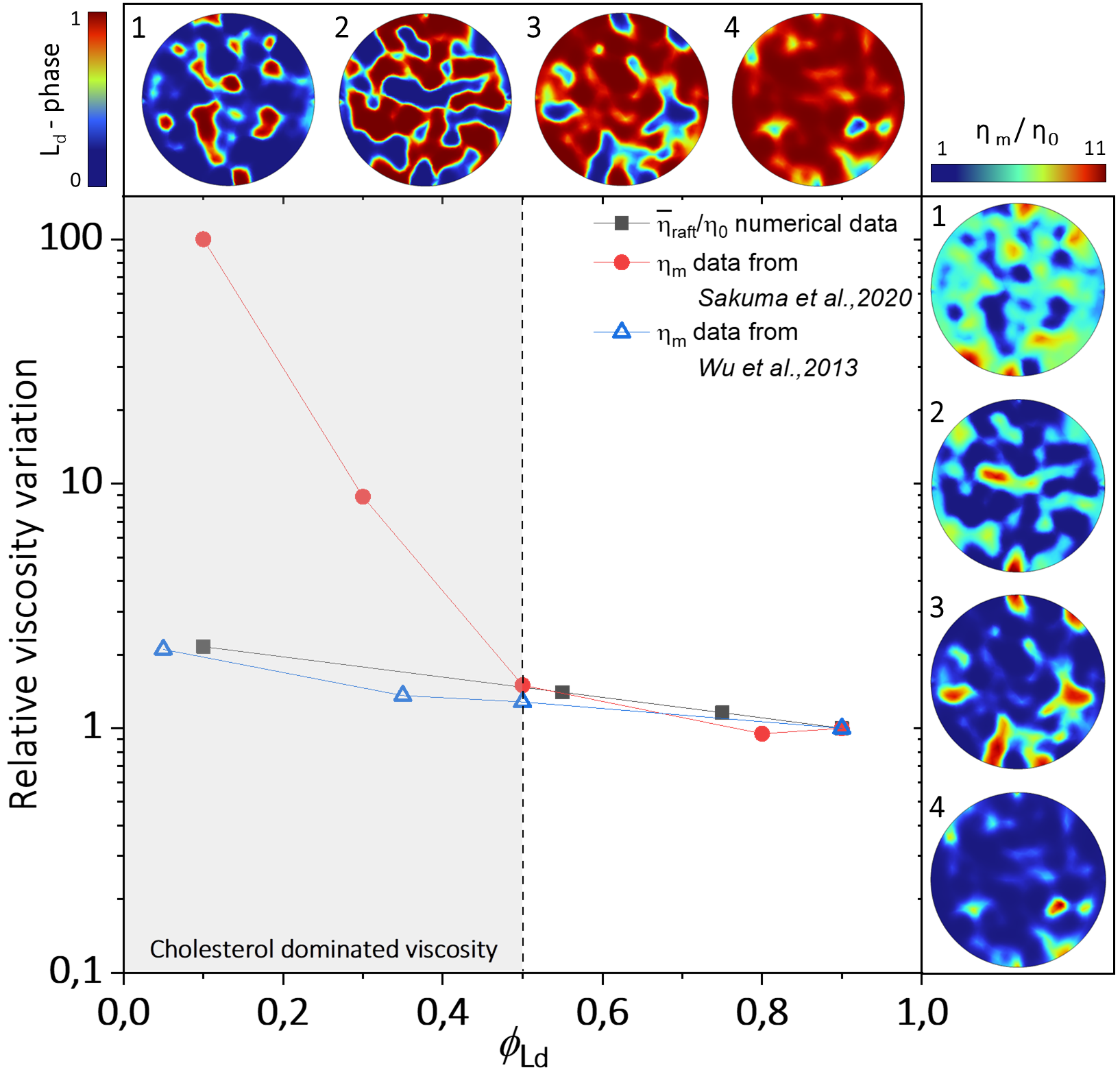}
    \caption{Numerical measured viscosities compared with experimental data adapted from \textit{Sakuma et al.}\cite{sakuma2020viscosity} and \textit{Wu et al.}\cite{wu2013molecular}. By assigning different spatial distributions in the ligand precipitation rate, in order to modulate the volume fraction of disordered domains, the model is capable to find consistent values with both the experimental findings in the range $0.5 \le \phi_{L_d} < 1.0$. Cholesterol rich membranes, $0 < \phi_{L_d} \le 0.5$, lead to variation in the measured viscosities that differ from the ones measured in absence of cholesterol percentages and the ones numerically found. Surface plots of disordered phase volume fractions are shown above and viscosity maps are visible on the right (adopted parameters $p_0=0.8MPa$ and $\tau=\tau_0$).
    }
    \label{exp}
\end{figure*}

To appraise these differences, we studied the viscosity behavior as a function of the volume fraction of the disordered phase $\phi_{L_d}$. This was done numerically by varying the amount of precipitating ligand, by so influencing the activation potential of the transmembrane proteins. As analyzed in Fig. \ref{exp}, the theoretical curve shows a two-fold viscosity ratio passing from a predominantly disordered phase to a domain mostly occupied by ordered clusters. 
These numerical outcomes have been put in direct comparison with two different sets of experimental data available in the literature. First, \textit{Sakuma et al.}\cite{sakuma2020viscosity} correlated the order parameter with the measured viscosity for different lipid systems. In such a case, the relative viscosity variations obtained from theoretical predictions well fit with these literature findings in the range $0.5 \le \phi_{L_d} < 1.0$. Below such an interval, i.e. for $0 < \phi_{L_d} \le 0.5$, the here presented model is far from capturing the experimental data obtained in \textit{Sakuma et al.}, as the reported values refer to lipid mixtures in which ordered and disordered phases coexist with a high cholesterol percentage. It is indeed confirmed that significant cholesterol percentages increase membrane viscosity\cite{doole2022cholesterol, cooper1978influence} and can impact on the change of membrane properties by chemically altering the lipid micro-environment. In the case at hand, for $0 < \phi_{L_d} \le 0.5$, these bilayers turn out to be rich in cholesterol content (about the $30\%$ more than the average ones) produced a different trend. 
In this sense, the lack of such species in the system represents a limitation, and more faithful results could be achieved by introducing a finer description of its role in the multi-physics model. More interestingly, the increase in viscosity predicted \textit{in silico} results that are remarkably compatible with additional literature findings over the entire range of phase order. In fact, the numerical curve is found to be in excellent agreement with data points derived from the experiments performed on giant unilamellar vesicles (GUVs) performed by \textit{Wu et al.}\cite{wu2013molecular}, in which lower Chol concentrations were employed. Noteworthy, they obtained a more gradual change of viscosity variation that increases to $2.1$ for ordered membrane configurations, so demonstrating the dynamic change of viscosity involved also in lipid rafts.

\section{Conclusions}
Following a recent theoretical formulation describing the mechanobiology of lipid membrane remodeling and raft formation carried out in\cite{bernard2023,CAROTENUTO2020103974}, the current study aims at investigating the dynamic visco-elastic response of plasma membranes to chemo-mechanical stimuli. Through \textit{in silico} analyses accounting for viscous-associated terms in the constitutive model, the multiphysics coupling between chemical events and mechanical adaptation highlights how the solid-fluid behavior of the bilayer evolves with the activity of the membrane. The evolved processes are strongly influenced by the dynamics of the transmembrane proteins activation and their interaction with the lipid medium. By considering both the cases of a Newtonian shear viscosity and a strain-sensitive viscosity, in this present paper we investigate the relationship between the reconfiguration of an initially inactive membrane micro-environment as a function of the competition between the internal viscous dissipation and the kinetics of phase transitions governing the emergence of lipid islands. 

Numerical outcomes allowed one to observe that the shear viscosity varies in phase-separated membranes resulting in higher values for ordered-phase domains, i.e. lipid rafts. Hence, this provides a mechanically-based explanation of a well-known phenomenon highlighted by a large number of biophysical studies by means of various experimental methods. 
The synergy between active protein regions and raft emergence leads the system to re-organize itself by creating thicker and more viscous domains. Also, sensitivity analyses revealed how the visco-elastic behavior is influenced by the intra-cellular pressure applied at the boundary. That alters the mechanical properties of the membrane, and the volume fraction of the liquid-disordered phase. Hence, our visco-elastic approach enriches the existing studies regulating the mechanisms on the lipid membrane's behavior. This could help to earn some insights in characterizing the role of lipid rafts in membrane mechanics and in mediating important cellular biochemical processes. 

By refining the modeling of species inter-specificity, one would have the opportunity to include some other agents influencing membrane dynamics in the analysis. This may allow one to enlarge the complex multi-species environment under exam, as well as to further enrich the membrane constitutive framework.
To this aim, the self-reconfiguration of lipids could be studied by considering non-convex terms in the elastic strain energy (see e.g.\cite{deseri2013stretching,CAROTENUTO2020103974} and reference cited therein). Moreover, enriched coupling terms may be considered in the model in order to have deeper insights into the influence of the mechanical stress on the interspecific dynamics. In fact, through \textit{ad hoc} mechanical feedback functions, it would be possible to better investigate the processes of cell mechano-sensing and mechano-trasduction, that inevitably involve the mediation of membrane selectivity during cell-environment communication.
Also, as emerged from the presented analyses, one of the main components that can be included to further refine and enrich the description of membrane visco-elastic adaptation could be the cholesterol. This has a direct responsibility for lipid rafts stabilization and bilayer lateral diffusion, GPCRs re-configuration and activity, besides its participation to determine the membrane effective properties. For this significant reason, this will be object of future investigations.

\section*{Appendix}
\subsection*{Strain-dependent tangent viscous properties}
\numberwithin{equation}{section}
\renewcommand{\theequation}{A.\arabic{equation}}
\setcounter{equation}{0}

Tangent viscosity has been evaluated by following a small-on-large approach\cite{carotenuto2019growth}. Except for the configurational factor $K_r$, starting from the second Piola-Kirchhoff stress: 
\begin{equation}
    \mathbf{S}=2\frac{\partial\mathcal{W}}{\partial\mathbf{C}}+\eta\,\mathbf{C}^{-1}\dot{\mathbf{C}}\mathbf{C}^{-1},
\end{equation}\label{A1}
a variation of this stress with respect to a certain finitely deformed configuration leads one to write $\mathbf{S}=\mathbf{S}_l+\delta\,\mathbf{S}$, where:
\begin{equation}\label{A2}
    \delta\,\mathbf{S}=\frac{\partial\mathbf{S}}{\partial\mathbf{C}}:\delta\,\mathbf{C}+\frac{\partial\mathbf{S}}{\partial\dot{\mathbf{C}}}:\delta\,\dot{\mathbf{C}}=\mathbb{C}_l:\delta\,\mathbf{C}+\mathbb{H}_l:\delta\,\dot{\mathbf{C}},
\end{equation}
in which $\mathbb{C}_l$ and $\mathbb{H}_l$ are elastic and viscous tangent material tensors, respectively. Under incompressibility, a push-forward of the Cauchy stress gives the following:
\begin{align}
    &\boldsymbol{\sigma}=\mathbf{F}\mathbf{S}\mathbf{F}^T=\delta\,\mathbf{F}\,\mathbf{F}_l\,\left(\mathbf{S}_l+\delta\,\mathbf{S}\right)\mathbf{F}_l^T\delta\,\mathbf{F}^T=\nonumber\\
    &\small{=\boldsymbol{\sigma}_l+\boldsymbol{\sigma}_l\,\mathbf{H}_\delta^T+\mathbf{H}_\delta\,\boldsymbol{\sigma}_l+\mathbf{F}_l\,\left(\mathbb{C}_l:\delta\,\mathbf{C}+\mathbb{H}_l:\delta\,\dot{\mathbf{C}}\right)\,\mathbf{F}_l^T},
\end{align}
where $\mathbf{H}_\delta$ is the displacement gradient associated to the small incremental deformation $\delta\,\mathbf{F}$. By exploiting the strain and strain-rate identities:
\begin{align}
    &\delta\,\mathbf{C}=\mathbf{C}-\mathbf{C}_l=\mathbf{F}_l^T\left[2\,\text{sym}(\mathbf{H_\delta})\right]\,\mathbf{F}_l=2\mathbf{F}_l^T\left[\boldsymbol{\varepsilon}_\delta\right]\,\mathbf{F}_l\nonumber,\\
    & \text{and} \nonumber\\
    &\small{\delta\,\dot{\mathbf{C}}=\dot{\mathbf{C}}-\dot{\mathbf{C}}_l=2\mathbf{F}_l^T\left[\mathbf{L}_l^T\boldsymbol{\varepsilon}_\delta+\boldsymbol{\varepsilon}_\delta\,\mathbf{L}_l\right]\mathbf{F}_l+2\mathbf{F}_l^T\dot{\boldsymbol{\varepsilon}}_\delta\mathbf{F}_l},
\end{align}
the updated Cauchy stress can be re-written as follows:
\begin{align}
    & \boldsymbol{\sigma}=\boldsymbol{\sigma}_l+ \left[\mathbf{I} \overline{\otimes} \boldsymbol{\sigma}_l+ \boldsymbol{\sigma}_l \underline{\otimes} \mathbf{I}\right]: \left[\boldsymbol{\varepsilon}_\delta+ \boldsymbol{\omega}_\delta\right] \nonumber \\ & + \left\{(\mathbf{F}_l\overline{\otimes}\mathbf{F}_l): \left[2\mathbb{C}_l\right]:(\mathbf{F}_l^T\overline{\otimes}\mathbf{F}_l^T) \right. \nonumber \\ & + \left.(\mathbf{F}_l\overline{\otimes}\mathbf{F}_l): \left[\mathbb{H}_l\right]:(\mathbf{F}_l^T\overline{\otimes}\mathbf{F}_l^T):(\mathbf{L}_l^T\underline{\overline{\otimes}}\mathbf{I}+\mathbf{I}\underline{\overline{\otimes}}\mathbf{L}_l)\right\}: \boldsymbol{\varepsilon}_\delta \nonumber \\ & + \left\{(\mathbf{F}_l\overline{\otimes}\mathbf{F}_l): \left[2\mathbb{H}_l\right]:(\mathbf{F}_l^T\overline{\otimes}\mathbf{F}_l^T) \right\}: \dot{\boldsymbol{\varepsilon}}_\delta,
\end{align}
where $\left[\mathbf{A}\overline{\otimes}\mathbf{B}\right]_{ijhk}=A_{ih}B_{jk}$, $\left[\mathbf{A}\underline{\otimes}\mathbf{B}\right]_{ijhk}=A_{ik}B_{jh}$ and $\left[\mathbf{A}\underline{\overline{\otimes}}\mathbf{B}\right]_{ijhk}=(A_{ih}B_{jk}+A_{ih}B_{jk})/2$. By focusing on the response to the incremental strain-rates, the tangent viscosity tensor can be evaluated as follows:
\begin{equation}\label{Hv}
    \mathbb{H}=\frac{\partial\,\boldsymbol{\sigma}}{\partial\,\dot{\boldsymbol{\varepsilon}}_\delta}=\left\{(\mathbf{F}_l\overline{\otimes}\mathbf{F}_l):\left[2\mathbb{H}_l\right]:(\mathbf{F}_l^T\overline{\otimes}\mathbf{F}_l^T)\right\}:\mathbb{S},\quad
\end{equation}
where $\mathbb{S}=(\mathbf{I}\underline{\overline{\otimes}}\mathbf{I})/2$ is the identity fourth-order tensor mapping symmetric  tensors. By virtue of \eqref{A1} and \eqref{A2}, and on account of constitutive expressions \eqref{sigma1} and \eqref{p1}, after some passages one has:
\begin{equation}
    \mathbb{H}=\eta(\mathbf{C})\,\left[\mathbb{S}-\text{sym}(\mathbf{I}\otimes(\textbf{e}_3\otimes\textbf{e}_3))\right].
\end{equation}
To measure the effective surface shear viscosity, a planar shear velocity $\mathbf{v}=v_1\,\mathbf{e}_1+v_2\,\mathbf{e}_2$ is imagined to be applied on a generic point of the upper membrane surface, by producing a shear deformation $\dot{\gamma}_s$ such that $dv=\dot{\gamma}_s\,dx_3$, or $dv_1=(\dot{\gamma}_s\,dx_3)\cos \theta_s$ and $dv_2=(\dot{\gamma}_s\,dx_3)\sin \theta_s$. Then, the corresponding strain rates are linked to the shear $\dot{\gamma}_s$ throught the relations:
\begin{align}
    &\dot{\varepsilon}_{13}=\frac{1}{2}\frac{\partial\,v_1}{\partial\,x_3}=\frac{1}{2}\dot{\gamma}_s\cos \theta_s,\nonumber\\
    &\text{and}\quad\nonumber\\
    &\dot{\varepsilon}_{23}=\frac{1}{2}\frac{\partial\,v_2}{\partial\,x_3}=\frac{1}{2}\dot{\gamma}_s\sin \theta_s.
\end{align}
Also, the associated testing shear stress is $\sigma_s=\sqrt{\sigma_{13}^2+\sigma_{23}^2}$. This implies that the effective (tangent) viscosity can be evaluated as follows:
\begin{align}
&\frac{\partial\, \sigma_s}{\partial\, \dot{\gamma}_s}=\nonumber\\
&=\small{\frac{1}{2\sigma_s}\left[2\sigma_{13}\frac{\partial\,\sigma_{13}}{\partial\,\dot{\varepsilon}_{13}}\,\frac{\partial\,\dot{\varepsilon}_{13}}{\partial\, \dot{\gamma}_s}+2\sigma_{23}\frac{\partial\,\sigma_{23}}{\partial\,\dot{\varepsilon}_{23}}\,\frac{\partial\,\dot{\varepsilon}_{23}}{\partial\, \dot{\gamma}_s}\right]=}\nonumber\\
&=\frac{1}{2}(\text{H}_{1313}\cos^2 \theta_s+\text{H}_{2323}\sin^2 \theta_s)=\eta(\mathbf{C}).
\end{align}
This equation is then used to express the viscosity variation $\overline{\eta}_{raft}$ observed on the raft domains.

\section*{Declarations}

\begin{itemize}
\item Funding

L.D., N.M.P. and C.B. gratefully acknowledge the Italian Ministry of Universities and Research (MUR) in the framework of the project DICAM-EXC, University of Trento, Departments of Excellence 2023-2027 (grant DM 230/2022).\\
A.R.C., L.D., M.F., and N.M.P. gratefully acknowledge the partial support from the Italian Ministry of Universities and Research (MUR) through the PON “Stream”-ARS01 01182.\\
L.D. gratefully acknowledges the partial support from (i) the grant PRIN-2022XLBLRX, (ii) the ERC-ADG-2021-101052956-BEYOND, (iii) partial support from the European Union through the ERC-CoG 2022, SFOAM, 101086644, (iv) the Italian Government through the 2023-2025 PNRR\_CN\_ICSC\_Spoke 7\_CUP E63C22000970007 grant, awarded to the University of Trento, Italy.\\
M.F. gratefully acknowledges the Italian Ministry for University and Research (MUR) for the grant FIT4ME-DROB (PNC0000007).\\
M.F. and N.M.P. acknowledge the financial support of the European Union -- Next Generation EU - Piano Nazionale di Ripresa e Resilienza (PNRR) -- MISSIONE 4 COMPONENTE 2, INVESTIMENTO N. 1.1, BANDO PRIN 2022 D.D. 104/ 02-02-2022 - (PRIN 2022 2022ATZCJN AMPHYBIA) CUP N. E53D23003040006.\\
N.M.P. and L.D. also gratefully acknowledge the partial support from the ERC through (1) FET Open “Boheme” grant no. 863179, (2) LIFE GREEN VULCAN LIFE19 ENV/IT/000213.\\
A.R.C. acknowledges the financial support of the European Union -- Next Generation EU - Piano Nazionale di Ripresa e Resilienza (PNRR) -- MISSIONE 4 COMPONENTE 2, INVESTIMENTO N. 1.1, BANDO PRIN 2022 PNRR D.D. 1409/14-09-2022 (PRIN 2022 PNRR no. P2022M3KKC MECHAVERSE ''MEchanics vs Cell competition: Hyperelasticity and Adaptation in Vascular Evolutionary Repair and Smart Endoprostheses'') CUP E53D23017310001.\\
This manuscript was also conducted under the auspices of the Italian Group of Theoretical Mechanics GNFM-INdAM of the \textit{National Institute of High Mathematics}.

\item Conflict of interest

The authors declare that they have no known competing financial interests or personal relationships that could have appeared to influence the work reported in this paper.

\item Ethics approval 

Not applicable
\item Consent to participate

Not applicable
\item Consent for publication

Not applicable
\item Availability of data and materials

Not applicable
\item Code availability 

All code for data analysis associated with the current submission is available under resonable requests.

\item Authors' contributions

All the Authors equally contributed to the study. Conceptualization:  L.D., M.F.,A.R.C.; Methodology: C.B., A.R.C., M.F.; Formal analysis and investigation: C.B., A.R.C.; Writing - original draft preparation: C.B., A.R.C.; Writing - review and editing: L.D., M.F., N.M.P.; Funding acquisition: L.D., M.F., N.M.P.; Supervision: A.R.C., L.D., M.F., N.M.P.
\end{itemize}

\bibliography{sn-bibliography}

\end{document}